\documentclass[traditabstract]{aa}
\usepackage{longtable,lscape,amssymb}
\usepackage{graphicx}
\usepackage{epsfig}
\usepackage[authoryear]{natbib}
\usepackage{natbib}

\bibpunct{(}{)}{;}{a}{}{,} 

\newcommand{\Ha}{{H$\alpha$}}

\newcommand{\Ht}{H$_2$}
\newcommand{\CIV}{C\,{\sc iv}}
\newcommand{\CII}{C\,{\sc ii}}
\newcommand{\MgII}{Mg\,{\sc ii}}
\newcommand{\NV}{N\,{\sc v}}

\newcommand{\pab}{{Pa$\beta$}}
\newcommand{\pag}{{Pa$\gamma$}}

\newcommand{\brg}{{Br$\gamma$}}

\newcommand{\Ll}{{$L_{\rm line}$}}

\newcommand{\Lacc}{{$L_{\rm acc}$}}
\newcommand{\Macc}{{$\dot{M}_{\rm acc}$}}

\newcommand{\Msun}{{$M_{\odot}$}}
\newcommand{\Lsun}{{$L_{\odot}$}}
\newcommand{\Rsun}{{$R_{\odot}$}}

\newcommand{\Mstar}{{$M_{\star}$}}
\newcommand{\Lstar}{{$L_{\star}$}}

\begin{document}

\title{HST spectra reveal accretion in MY Lupi \thanks{
Based on observations made with the NASA/ESA Hubble Space Telescope, obtained from the data archive at the 
Space Telescope Science Institute. STScI is operated by the Association of Universities for Research in 
Astronomy, Inc. under NASA contract NAS 5-26555. Also based on  observations collected at the European 
Southern Observatory at Paranal, under programs  095.C-0134(A), and archive data of programmes 
093.C-0476(A) and 093.C-0506(A).
 }
}

\author{
       J.M. Alcal\'a\inst{1} 
  \and C.F. Manara\inst{2}
  \and K. France\inst{3}
  \and C.P. Schneider\inst{4}
  \and N. Arulanantham\inst{3}
  \and A. Miotello\inst{2}
  \and H.M. G\"unther\inst{5} 
  \and A. Brown\inst{6}
}


\institute{ 
      INAF-Osservatorio Astronomico di Capodimonte, via Moiariello 16, 80131 Napoli, Italy
  \and European Southern Observatory, Karl-Schwarzschild-Str. 2, D-85748 Garching bei Munchen, Germany
  \and Laboratory for Atmospheric and Space Physics, University of Colorado, 392 UCB, Boulder, CO 80303, USA
  \and Hamburger Sternwarte, Gojenbergsweg 112, D-21029 Hamburg, Germany
  \and MIT, Kavli Institute for Astrophysics and Space Research, 77 Massachusetts Avenue, Cambridge, MA 02139, USA
  \and Center for Astrophysics and Space Astronomy, University of Colorado, Boulder, CO 80309-0389, USA
}

\date{Received ; accepted  }

\abstract{ 
The mass accretion rate is a crucial parameter for the study of the evolution of accretion discs around young 
low-mass stellar and substellar objects (YSOs). We revisit  the case of  MY\,Lup, an object 
where VLT/X-Shooter data suggested a negligible mass accretion rate, and show it to be accreting on a level 
similar to other Class~II YSOs in Lupus based on Hubble Space Telescope (HST)  observations. 
In our HST-Cosmic Origins Spectrograph (HST-COS) and -Space Telescope Imaging Spectrograph (HST-STIS) 
spectra, we find many emission lines, as well as substantial far-ultraviolet (FUV) continuum excess emission, 
which can be ascribed to active accretion.
The total luminosity of the \CIV~$\lambda$1549\,\AA ~doublet is 4.1$\times10^{-4}$\,\Lsun. Using scalings 
between accretion luminosity, \Lacc, and \CIV ~luminosity from the literature, we derive \Lacc $\sim$2$\times10^{-1}$\Lsun, 
which is more than an order of magnitude higher than the upper limit estimated from the X-Shooter observations. 
We discuss possible reasons for the X-Shooter-HST discrepancy, the most plausible being that the low contrast 
between the continuum excess emission and the photospheric$+$chromospheric emission at optical wavelengths
in MY\,Lup hampered detection of excess emission.
The luminosity of the FUV continuum and \CIV ~lines, strong \Ht ~fluorescence, and a "1600 A Bump" place MY\,Lup 
in the class of accreting objects with gas-rich discs.  So far, MY\,Lup is the only peculiar case in which a 
significant difference between the HST and X-Shooter \Macc ~estimates  exists that is not ascribable to variability.
The mass accretion rate inferred from the revisited \Lacc ~estimate is \Macc ~$\sim$ 1($^{+1.5}_{-0.5}$)$\times10^{-8}$\,\Msun yr$^{-1}$. 
This value is consistent with the typical value derived for accreting YSOs of similar mass in Lupus 
and points to less clearing of the inner disc than indicated by near- and mid-infrared observations. 
 This is confirmed by Atacama Large Millimeter Array (ALMA) data, which show that the gaps and rings seen 
 in the sub-millimetre are relatively shallow.
}

\keywords{Stars: pre-main sequence, low-mass -- Accretion, accretion disks -- protoplanetary disks - single objects: MY\,Lup}

\titlerunning{Accretion in MY\,Lup}
\authorrunning{Alcal\'a et al.}
\maketitle

\section{Introduction} 
\label{intro}

A key issue in the study of planet formation is to explain how optically thick accretion discs surrounding 
the youngest low-mass (\Mstar$\lesssim$2.0\Msun) stars evolve into optically thin debris discs. 
The mass accretion rate onto a young star, \Macc, is a crucial parameter for the study of the evolution of 
accretion discs around young low-mass stellar and substellar objects (YSOs), because it sets important 
constraints for disc-evolution models \citep{hartmann98, hartmann16} and disc-clearing mechanisms 
\citep[][and references therein]{alexander14,ercolanopascucci17}, and is a key quantity for the studies of 
pre-main sequence (PMS) stellar evolution and planet formation \citep[][]{morbidelliraymond16}. 

Transition discs (TDs) are protoplanetary discs that show evidence of inner holes and gaps, 
as observed in millimetre interferometric observations \citep[e.g.][]{andrews11,vandermarel18,andrews18} 
and in the dip of the mid-infrared spectral energy distribution (MIR; SED) \citep[e.g.][]{merin08}.
Low-mass PMS stars with transitional discs accreting at very low rates are likely in the final stages 
of inner disc evolution, and have probably already formed proto-planets \citep[][]{owen_clarke12}. 
Therefore, identifying and investigating such slow accretors may help us to understand planet formation. 
However, measurements of low \Macc ~are challenging. In practice,   \Macc ~can be derived from the energy released in 
the accretion shock \citep[accretion luminosity \Lacc, see][]{gullbring98, hart98} given the stellar 
properties. 
Observationally, this requires measurements of excess flux in continuum and lines with respect to
similar non-accreting stars.
 Such measurements are best performed at UV wavelengths ($\lambda$ $\lesssim$ 4000\,\AA), 
with the Balmer continuum excess emission and the Balmer jump being more easily seen in the spectra of 
late-type ($>$~K5) YSOs than in the early types due to the higher contrast between photospheric emission 
and continuum emission \citep[][]{HH08}. Weak accretion, in general, is not easily detectable in the region 
of the Balmer jump. Some previous studies attempted \Macc ~measurements in TDs using other tracers.
In the case of CVSO\,224, a 10\,Myr old PMS star with a TD, \citet[][]{espaillat08} derived 
\Macc$=$7$\times10^{-11}$\,\Msun yr$^{-1}$ based on modelling of the \Ha ~emission line observed 
at high spectral resolution. Other estimates of \Macc ~in samples of YSOs with TDs have been
provided using the width, W\Ha(10\%), of the \Ha ~line measured at the 10\% of the line peak in 
intermediate-resolution spectra \citep[][]{merin10, cieza10}. However, such estimates heavily depend on 
rather uncertain scaling relations between the line width and \Macc ~\citep[see discussion in][]{alcala14}.

We measured the UV excess continuum emission for a ~90\% complete sample of YSOs in the Lupus star-forming 
region  \citep{alcala14, alcala17} using the VLT/X-Shooter spectrograph \citep{vernet11}. 
This dataset allowed us to derive the stellar and accretion parameters in a self-consistent and 
homogeneous way \citep[see also][for the methods]{manara13}.
In the sample of 81 systems, 12 are YSOs with TDs based on the dip of their MIR SEDs \citep[][]{merin08, romero12, bustamante15}. 
Among these, we identified 5 objects (Lup\,607, MY~Lup, Sz65, Sz68, and SST\,c2dJ160830.7-382827) in which the UVB excess 
emission, ascribable to accretion, is barely evident. The analysis of the emission lines in the 
X-Shooter spectra of these objects showed that their excess emission is close to the chromospheric 
noise level as defined by \citet[][see also Ingleby et al. 2011]{manara13a, manara17}. 
We therefore classified them as weak or non-accretors.
Two of these (MY~Lup and SST\,c2dJ160830.7-382827) were previously classified as transitional 
discs and their spectral type is earlier than K5. It was thus unclear whether the apparent low level 
of accretion in these objects is real or due to a very low contrast between the continuum excess 
emission and the photospheric$+$chromospheric emission, which prevented a reliable assessment of 
accretion. 

Observations in the far-ultraviolet (FUV) are particularly well suited for studies of accretion.
Previous observations in the FUV have shown strong accretion-related emission features. For instance,
strong emission of CO \citep[e.g.][]{france11, schindhelm12} and \Ht ~ascribed to gas in the inner regions 
($r\lesssim$10\,AU) of the disc of accreting YSOs 
\citep[][and references therein]{johns-krull00, herczeg02, calvet04, france11, ingleby11, yang12, 
france12, hoadley15,france17, arulanantham18}.
This is of particular importance for the studies of accretion in YSOs because these molecules trace 
the gas in the inner disc regions, indicating that molecular material persists within a few astronomical 
units of the star. 
In this work, we use UV-Hubble Space Telescope (HST) spectra to unambiguously confirm and investigate the accretion properties
of MY\,Lup. In a forthcoming paper \citep[][]{Arulanantham_prep}, an extended set of the 
same data is used to study the structure and distribution of the gas in the inner disc regions. 
In Section~\ref{prevobs} we present an overview of MY\,Lup and previous spectroscopic observations
used to provide intial estimates of mass accretion rate, \Macc . In Section~\ref{obs} we report the HST 
observational data, their processing, and the main spectroscopic features in the near-ultraviolet (NUV)
and FUV. 
In Section~\ref{datanalysis} we describe the analysis of the HST spectra and report the results 
on the accretion properties of MY\,Lup. In Section~\ref{discus} we discuss the implications of 
the assessment of accretion in MY\,Lup and its disc. Finally, we present our conclusions 
in Section~\ref{conclu}.

\section{The target} 
\label{prevobs}
 
MY\,Lup was first classified as a T~Tauri star by \citet[][]{gregoriohetem92} based on the presence 
of the \Ha ~emission line and the Li\,{\sc i}$\lambda$6708\AA ~absorption line in an intermediate-
resolution spectrum.
It is a K0-type star with a mass of 1.1\,\Msun ~\citep[][]{alcala17, frasca17} in the Lupus\,IV 
cloud \citep[][]{comeron08}, surrounded by a highly inclined ($\sim$73$^\circ$) disc  
\citep[][]{ansdell16, vandermarel18, andrews18, huang18}.  \citet[][]{romero12}
classified MY\,Lup as a planet-forming TD disc candidate, claiming that its SED could be 
explained by a discontinuity in the grain size distribution rather than an inner opacity hole.
Recent high-contrast high-resolution imaging in the near-infrared (NIR) with the 
Spectro-Polarimetric High-contrast Exoplanet REsearch (SPHERE) instrument by 
\citet[][]{avenhaus18} confirmed a high inclination of $\sim$77$^\circ$ of the dusty disc. 
The SPHERE observation 
also revealed a flared-truncated disc structure with multiple rings on the surface, and 
 the Disk Substructures at High Angular Resolution Project  \citep[DSHARP][]{andrews18} 
 also detected several annular substructures at 8, 20, 30, and 40\,au from the central star 
\citep[][]{huang18}, with no evidence of a large inner cavity.

The age of the relatively extincted \citep[A$_{\rm V}$=1.3\,mag, ][]{alcala17} star has been estimated 
at about 17\,Myr  \citep[c.f.][]{alcala17, frasca17}, that is, a factor of six older than typical YSOs in Lupus, 
suggesting that the object may be more evolved than other Lupus YSOs or that the highly inclined 
disc occults part of the stellar light, making the object sub-luminous on the HR diagram. 
A tentative stellar rotational period of 2.5\,days has been reported in \citet[][]{batalha98}.

\subsection{Previous spectroscopy and \Macc ~estimates}

Previous VLT spectroscopy of MY\,Lup includes two X-Shooter spectra, one by us \citep[reported in][]{alcala17} 
and the other one obtained within ESO programme 095.C-0134 (PI. Caceres). In addition, a UVES spectrum 
was acquired within ESO programme 093.C-0476 (PI. Canovas). The latter was performed 6 days after the 
Caceres X-Shooter observation, while our X-Shooter spectrum was acquired about 1 year later. 
The ESO Phase-3 reduced data of both the Caceres X-Shooter and Canovas-UVES spectra are used here 
for comparison purposes.

Except for \Ha ~and the Ca\,{\sc ii} H \& K and IRT emission lines, which appear superimposed
on the corresponding photospheric absorption, no other emission lines are clearly seen in 
the VLT spectra of MY\,Lup. In both the X-Shooter spectra the He\,{\sc i} $\lambda$1082.9\,nm
appears as a very strong absorption with a possible but barely detectable emission, resembling 
a P-Cyg profile. 
When subtracting a photospheric template, still no other emission lines are evident; only
the \pag, \pab, and \brg ~lines are barely seen, but detected  \citep[the fluxes are given in][]{frasca17}. 
A possible explanation for why we see the lines in the red but no other lines in the blue may be 
that the highly inclined disc occults some of the emitting inner regions, with the lines bluer 
than \Ha ~being more extincted and therefore remaining undetected.
 
\begin{figure}[h]
\resizebox{1.0\hsize}{!}{\includegraphics[]{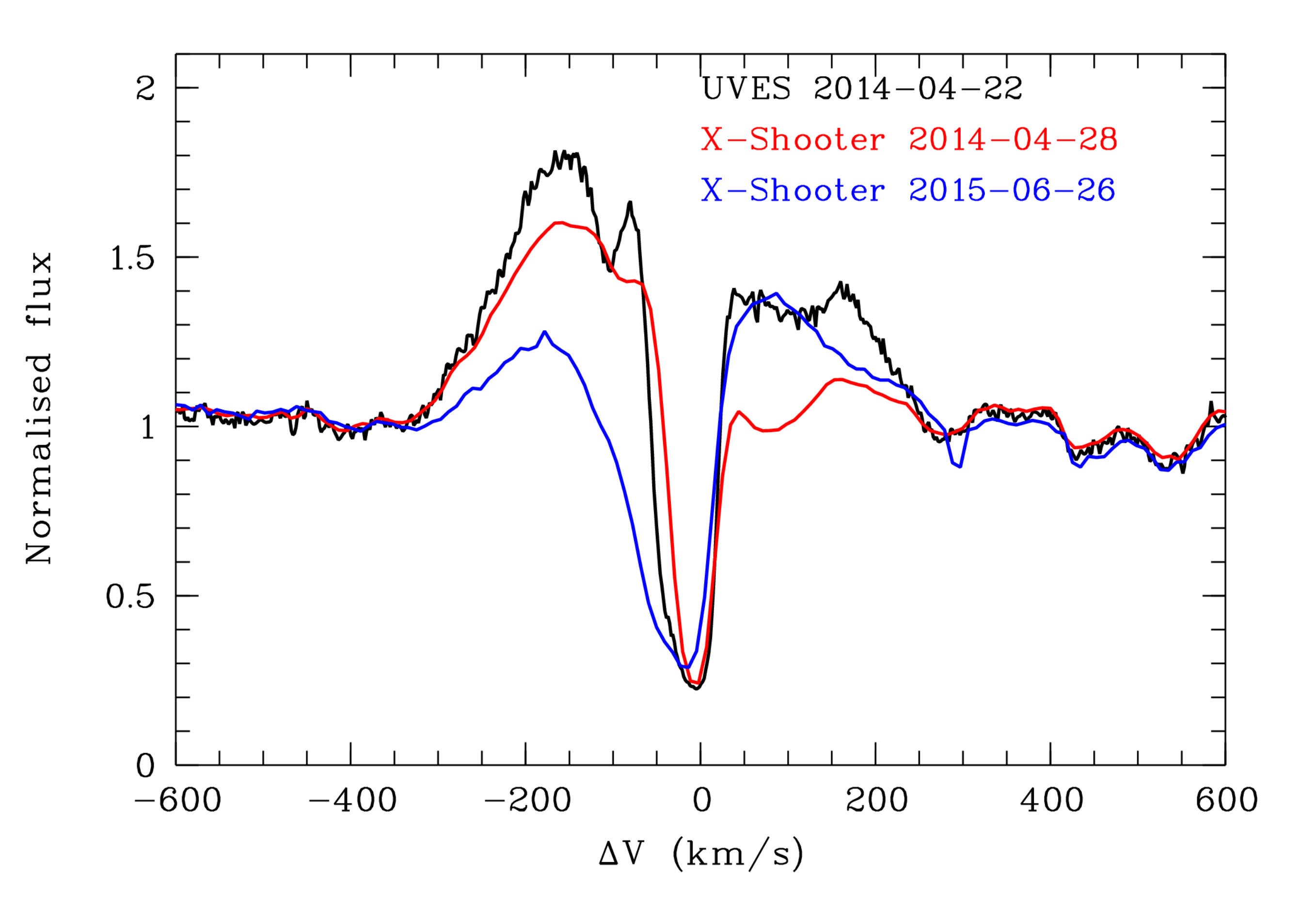}}
\caption{\Ha ~ line profile of MY\,Lup observed in three epochs as labelled. 
        The spectra are normalised to unity and plotted in velocity scale. 
    \label{Ha_var}}
\end{figure}

We highlight the fact that the \Ha ~line shows a complex profile, several hundred km~s$^{-1}$ wide 
(c.f. Figure~\ref{Ha_var}). 
The width at 10\% of the line peak measured in the UVES spectrum is $\sim$500\,km\,s$^{-1}$. This value
is consistent with the one estimated by \citet[][]{romero12} using other spectra.
We also note that the X-Shooter spectrum we used in \citet[][]{alcala17} has the weakest \Ha ~emission.

Variability of the \Ha ~line profile is observed on a timescale of 1 year, but some variations 
are detected within a few days (see Figure~\ref{Ha_var}). The type of profile is consistent with 
the model predictions by \citet[][see their Figure~13]{kurosawa06} for a high-inclination angle
of accreting sources.
In those models, the absorption feature becomes stronger as the inclination increases, mainly 
because of the geometrical configuration. A high inclination is in agreement with the disc 
inclination estimated from the Atacama Millimeter Array (ALMA) millimetre \citep[][]{ansdell16, vandermarel18} and SPHERE 
NIR images \citep[][]{avenhaus18}, and is consistent with the relatively high stellar 
rotation rate of $\sim$29\,km\,s$^{-1}$ \citep[c.f.][]{frasca17}. 

The U-excess continuum emission in MY\,Lup is not evident \citep[see Figure~E.8 in][]{alcala17} 
when using our methods \citep[see][]{manara13, alcala14, alcala17} to derive the accretion luminosity, 
\Lacc. 
Based on that analysis, which basically measures the Balmer continuum excess emission with 
respect to the photospheric one, including also the chromospheric level from a non-accreting template, 
we estimated an upper limit on \Lacc ~of 0.005\,\Lsun, which means an upper limit on 
\Macc $<$ $2.24\times10^{-10}$\,\Msun yr$^{-1}$, placing  MY\,Lup very 
close to the chromospheric noise level, and more than an order of magnitude below the typical 
\Macc ~value for YSOs of similar mass in the Lupus star forming region. We have also verified that the 
estimates of the limit on \Lacc ~ derived from the Balmer continuum and the few emission lines detected
are consistent \citep[see][]{alcala17}. According to the \Macc--W\Ha(10\%) scatter plot 
for the Lupus sample shown in Figure~9 of \citet[][]{alcala14}, a W\Ha(10\%)$\sim$500\,km\,s$^{-1}$ 
measured in the UVES spectrum would suggest $\log$\Macc$\approx$$-8.6$ for MY\,Lup, 
but the spread of \Macc~versus~W\Ha(10\%) is very large. Moreover, in the X-Shooter spectrum 
used in \citet[][]{alcala17}, the \Ha ~line was not in emission as much as in the UVES and the 
other X-Shooter spectrum (see Figure~\ref{Ha_var}). 
All this may lead to the conclusion that MY\,Lup is not accreting.
This has been claimed to be the case for several transition discs, and in principle would be in 
line with the expectation from photoevaporation models \citep[][and references therein]{ercolano17, ercolanopascucci17}. 
However, it may be at odds with the detection of significant excess emission in the MIR and ALMA millimetre bands
 \citep[see SED in Figure~2 of][and in Figure~1 of Andrews et al. 2018]{vandermarel18}.
From radiative transfer modelling, which accounts for optical depth and disc inclination, \citet[][]{miotello17} 
find a total dust mass of $\sim$173\,M$_{\rm Earth}$.

\section{HST data} 
\label{obs}

\subsection{Observations and data processing}
 
The HST data used in this paper were acquired with the Cosmic Origins Spectrograph (COS) and
the Space Telescope Imaging Spectrograph (STIS) onboard the HST as part of Cycle~24 General Observer 
program (PID 14604; PI: C.F. Manara).
The data were acquired in September 8, 2017 (UT start 01:35:54; UT end 10:07:17), which is more 
than two years after our X-Shooter observation of June 26, 2015.

The data were collected in three different observing modes of HST-COS with the  
gratings G140L, G130M, and G160M, centred at $\lambda$1280, $\lambda$1291, and $\lambda$1577, providing 
a resolving power $\sim$1500, $\sim$16000 and $\sim$16000, respectively \citep[see][]{green12}, as well as 
in two observing modes of HST-STIS using gratings G430L  and G230L, centred at $\lambda$4300 and 
$\lambda$2375, respectively,   both with a resolving power of $\sim$1000 \citep[see][and references therein]{woodgate98}.

The G160M exposure time was 2592 sec through the primary science aperture, with observations 
split between the 1577 and 1611 cenwave settings to mitigate fixed pattern noise and provide continuous spectral 
coverage over the \CIV, \Ht, and FUV continuum emitting wavelength regions.  The velocity and flux calibration 
of the G160M spectra are $\pm$ 15 km s$^{-1}$ and 5\% absolute, respectively.  The G130M and G140L spectra will 
be presented in a future work (Arulanantham et al. in prep).

The data processing has been done in the same way as described in \citet[][]{arulanantham18}.
Briefly, individual flux-calibrated spectra generated by the COS pipeline are coadded using a 
cross-correlation algorithm \citep[][]{danforth10} that has been optimised for emission line sources 
\citep[e.g.][]{france12}. The data from different grating settings are then joined at the spectral 
overlap regions, with interpolated fluxes used to connect the observations from different gratings.   

\begin{figure}[h]
\resizebox{1.0\hsize}{!}{\includegraphics[]{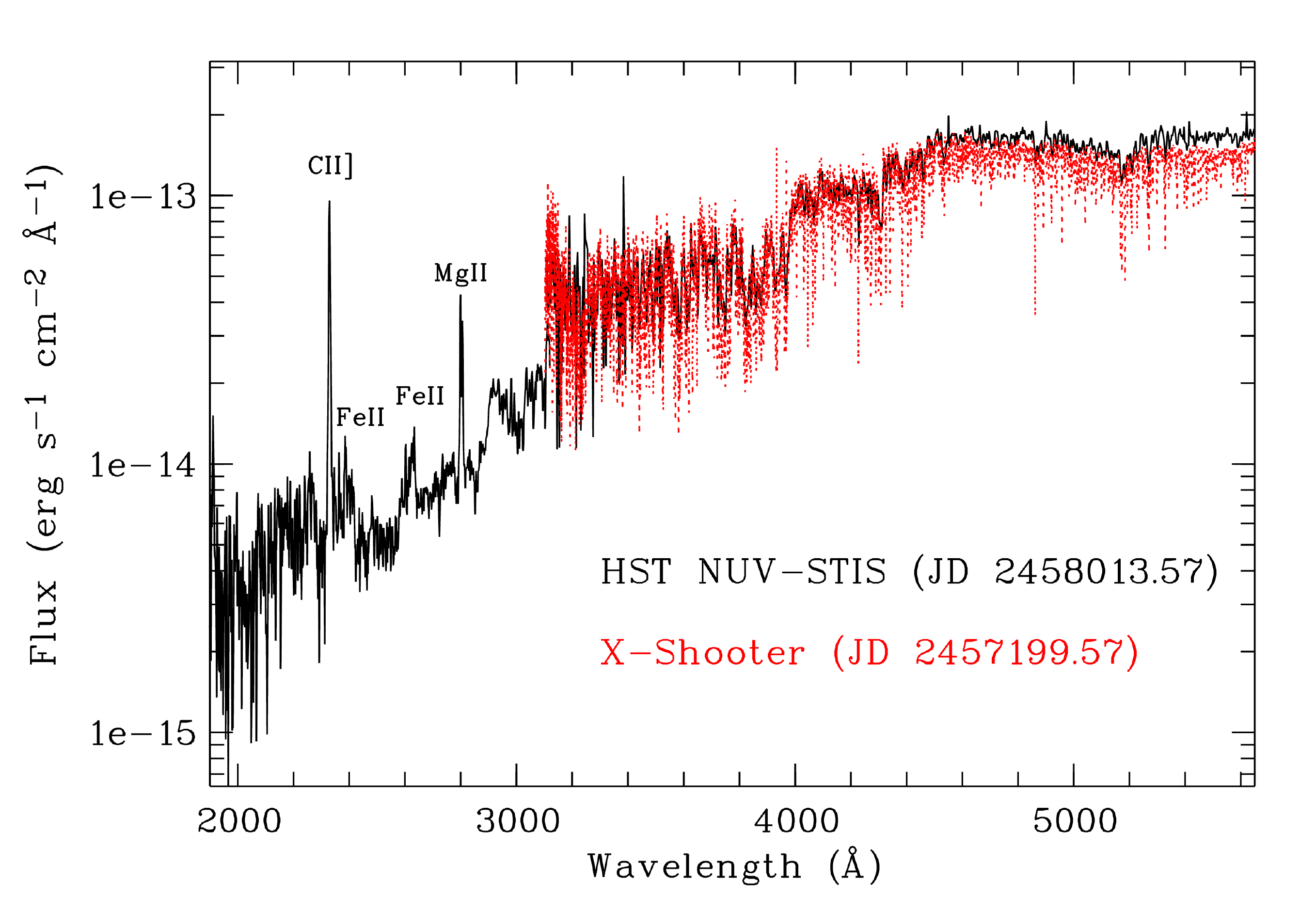}}
\caption{Hubble Space Telescope NUV and STIS spectra of MY\,Lup are shown as the black continuous line. The red dotted
line represents the X-Shooter spectrum acquired more than 2 years earlier as indicated by the JD in the 
labels. The spectra are corrected for reddening as described in the text.
   \label{spec_uv_x_shoot}}
\end{figure}

\subsection{Ultraviolet HST spectra}

Figure~\ref{spec_uv_x_shoot} shows the HST STIS spectra. Our X-Shooter spectrum is overplotted as the red 
dotted line for comparison purposes. The spectra are corrected for reddening using A$_{\rm V}=$1.3\,mag 
and adopting the extinction law by \citet[][]{cardelli89} with R$_{\rm V}=$3.1 
\citep[however, see discussion on reddening in][]{calvet04}. We note that the fluxes of the 
HST-optical and X-Shooter spectra are very consistent, despite being acquired more than 2 years apart. 
The very small slope  decrease in the reddest spectral range can be explained by an extinction change 
$\Delta$A$_{\rm V}$~$<$~0.2\,mag. However, this is within the uncertainties of the X-shooter analysis.
The other X-Shooter spectrum is also consistent with the HST data.
Strong emission in the \CII] $\lambda$2325 and \MgII ~$\lambda$2800 lines is also detected in the NUV spectrum. 
The latter line is  related to accretion, but has also been observed in non-accreting stars, while the former 
is much stronger in accretors than in non-accretors \citep[][]{ingleby11} and has been observed in other actively 
accreting sources \citep[][]{calvet04, gomezdecastro05, ingleby11}. 
The tentative presence of the [O\,{\sc i}]~$\lambda$6300 ~forbidden line may in principle lead to the
idea that the semi-forbidden  \CII] $\lambda$2325 line in MY\,Lup might form in a less dense environment 
than in the accretion flows, possibly in an outflow. However, only an upper limit for the flux of the 
[O\,{\sc i}] line could be derived in \citet[][]{nisini18}. Moreover, the outflow interpretation for the
\CII] $\lambda$2325 line is for stars that have strong outflows \citep[][]{gomezdecastro05}. Therefore, 
the \CII] $\lambda$2325 emission line in MY\,Lup is mostly formed in the accretion flows.

\begin{figure}[h]
\resizebox{1.0\hsize}{!}{\includegraphics[bb=50 70 700 530]{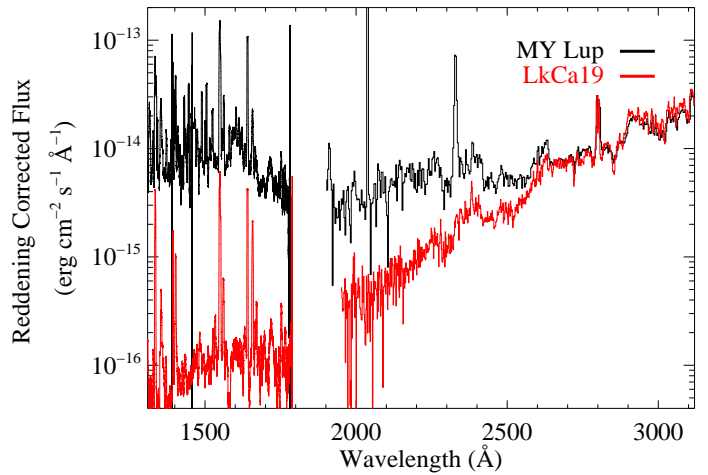}}
\caption{Reddening corrected FUV spectrum, and part of the NUV spectrum, of MY\,Lup are shown 
        with the black lines. The spectra of the T~Tauri star LkCa\,19 are overplotted with red lines. 
        The FUV continuum, \Ht ~fluorescence, and 1600\AA ~Bump, characteristic of accreting objects 
        \citep[][]{france17}, are observed in the spectrum of MY Lup.
   \label{MYLup_LkCa19}}
\end{figure}

In Figure~\ref{MYLup_LkCa19} the reddening corrected FUV spectrum, and part of the NUV spectrum, of MY\,Lup 
are shown. The spectra of the non-accreting YSO LkCa~19 are also overplotted as a red line for comparison. 
This star is also of K0-type, has a mass of 1.36\,\Msun, similar within errors to MY\,Lup, and extinction 
A$_V=0$\,mag \citep[see][]{kenyonhartmann95,kraushillenbrand09} \footnote{We note that \citet[][]{HH14} adopt
also a low value of A$_V=0.5$\,mag.}. Its projected rotational 
velocity was measured at 18.6\,km~s$^{-1}$ \citep[][]{hartmann87} and a rotational period of 2.2\,days has 
been reported in \citet[][]{clarke00}. It is confirmed that there are no signs of accretion in this star 
and therefore it can be used as a good comparison template to asses the accretion activity 
in MY\,Lup. The spectra of LkCa~19 \citep[][]{ardila13} were smoothed to $\sim$3\AA ~spectral 
resolution to be comparable to the MY\,Lup spectrum. 

\begin{figure}[h]
\resizebox{1.0\hsize}{!}{\includegraphics[angle=-90, bb=54 54 558 738]{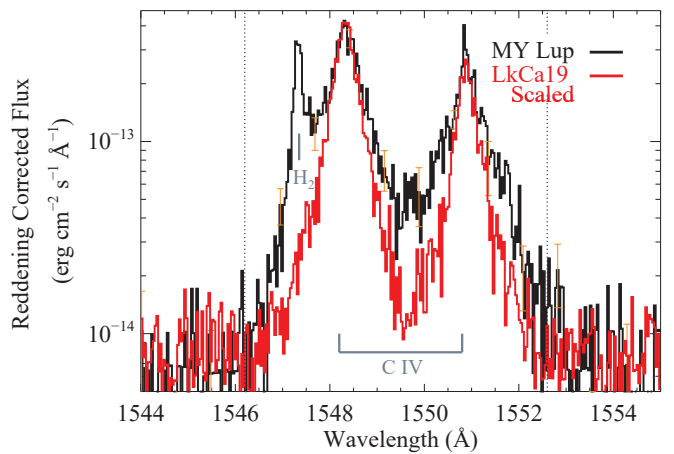}}
\caption{Far-ultraviolet spectrum of MY\,Lup in the region of the \Ht ~and \CIV ~doublet is shown with the 
black continuous line. The spectrum of the non-accreting young star LkCa\,19 is overplotted 
as a red line. 
   \label{MYLup_CIV_vs_RECX1}}
\end{figure}

A significant UV continuum excess emission and many emission lines tracing the inner disc (\Ht ~and CO) 
and star--disc interaction  region (\NV, \CIV, etc)  are detected in MY\,Lup. 
In this work, we focus on a few lines that allow us to investigate the accretion properties,
in particular, the \CIV ~$\lambda$1549 doublet, and the \Ht ~$\lambda$1547 and \CII] $\lambda$2325 lines. 
In a forthcoming paper \citep[][]{Arulanantham_prep}, a more detailed analysis of the complete set of 
detected lines will be presented. 

The FUV spectrum of MY\,Lup shows both the \CIV ~$\lambda$1549 doublet and dozens of \Ht ~lines
including the \Ht ~$\lambda$1547 ~molecular emission (see Figure~\ref{MYLup_CIV_vs_RECX1}). 
The \Ht ~lines are normally ascribed to gas-rich inner discs around accreting YSOs
\citep[see][and references therein]{france12, ingleby11, ardila13, hoadley15}. \Ht ~ fluorescence is observed 
in all accreting protostars observed with HST to date, while no stars without dust discs displaying 
\Ht ~emission have been discovered \citep[][]{france12}.  

The spectrum of the non-accreting YSO LkCa~19 is also overplotted in Figure~\ref{MYLup_CIV_vs_RECX1} 
as a red line for comparison. This comparison shows that the profiles of the \CIV ~($\lambda\lambda$1548.2, 1550.8) 
lines in MY\,Lup are broader than those of the non-accreting star, leading to the conclusion that 
MY\,Lup is accreting.

In summary, the presence of  \Ht ~lines, the higher level of FUV continuum flux with respect to that of 
the non-accreting star, and the presence of the strong \CII] $\lambda$2325 line, as well as the \Ha ~profile 
and its variability, all strongly suggest that MY\,Lup is an actively accreting object.

\section{Analysis of HST-UV spectra}
\label{datanalysis}

\subsection{Ultraviolet continuum emission}
\label{continuum_emiss}

We can use the slab model derived for MY\,Lup in \citet[][]{alcala17} and the spectrum of the non-accreting 
star to investigate by how much the UV excess emission is eventually underestimated. 

\begin{figure}[h]
\resizebox{1.0\hsize}{!}{\includegraphics[angle=-90, bb=0 60 550 750]{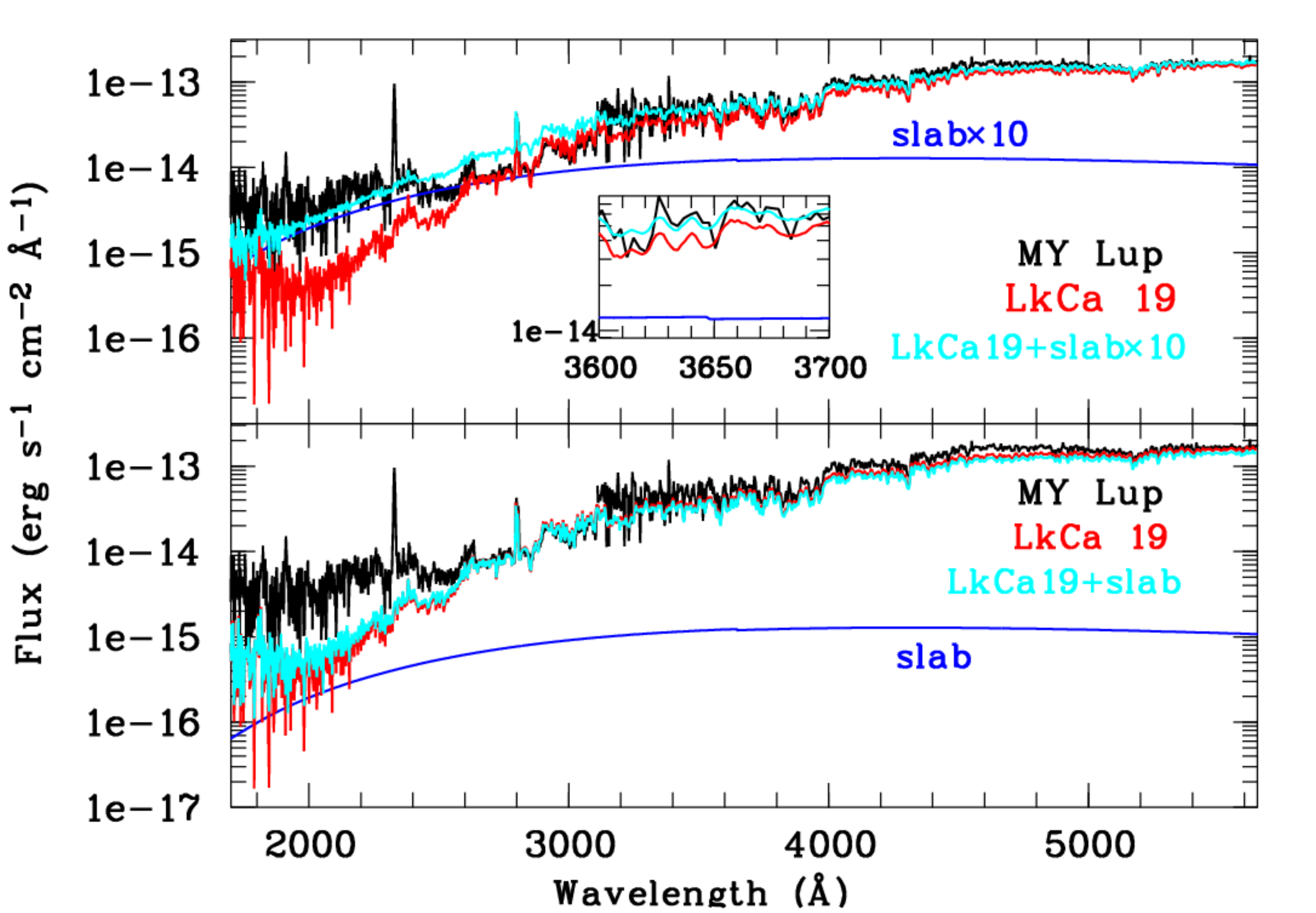}}
\caption{Lower panel: Extinction corrected HST NUV and STIS spectra of MY\,Lup (black) and LkCa\,19 (red). 
  The slab model derived in \citet[][]{alcala17} is also shown (blue). The cyan spectrum shows the sum of 
  LkCa\,19 plus the slab. Upper panel: Same as in lower panel, but with the slab model multiplied by ten.
  We note the tiny Balmer Jump shown in the inset. 
   \label{slab_fit}}
\end{figure}

The lower panel of Figure~\ref{slab_fit} shows how the slab model derived from the X-Shooter data
in our previous analysis, plus the spectrum of the non-accreting star, fail to match the level 
of NUV continuum flux from MY\,Lup, albeit still with a relative good match in optical bands. 
At about 2000\AA ~ this model yields at least a factor seven less flux than observed, and as shown in 
the upper panel of the same figure, a slab at least a factor ten brighter would be required to achieve 
the continuum flux in the NUV. This would imply an accretion luminosity of more than an order of magnitude 
higher than our estimates based on X-Shooter and is consistent with the results of the analysis of 
the \CIV ~emission lines, as is described in the following sections.
While in our previous analysis we used HBC\,407 as a non-accreting template, also a K0-type star 
\citep[][]{HH14}, the optical spectrum is also well fitted when using LkCa\,19 as a template.

As can be seen from the inset in Figure~\ref{slab_fit}, the X-Shooter spectrum shows no significant 
Balmer jump, which may be reconciled by fitting the spectrum with an almost optically thick slab. However, 
the fit deviates from the photospheric features at short wavelengths, that is, the observed NUV spectrum 
is fainter than the slab fit over some wavelengths ($\sim$2400\,\AA$-$3000\,\AA). This may be because 
the fit is very sensitive to the depth of photospheric features and therefore to any mismatch between 
the spectral features of MY\,Lup and those of the templates.

\subsection{The \CIV ~$\lambda$1549 doublet }
The \CIV ~ $\lambda$1549 line emission in YSOs has been shown to correlate with the accretion luminosity, 
and therefore with the mass accretion rate, indicating that it is most likely originated in the accretion shock and 
flows \citep[][]{johns-krull00, ingleby11, ardila13, france14}. 
To quantify how much of the \CIV ~$\lambda$1549 flux in MY\,Lup can be attributed to excess accretion emission, 
one should compare the fluxes to those of the non-accreting star, scaled to match the peak fluxes of the narrow 
components, because these are likely dominated by the magnetic activity in the transition region of the
 protostellar atmosphere.

\begin{figure}[h]
\resizebox{1.0\hsize}{!}{\includegraphics[bb=90 80 700 550]{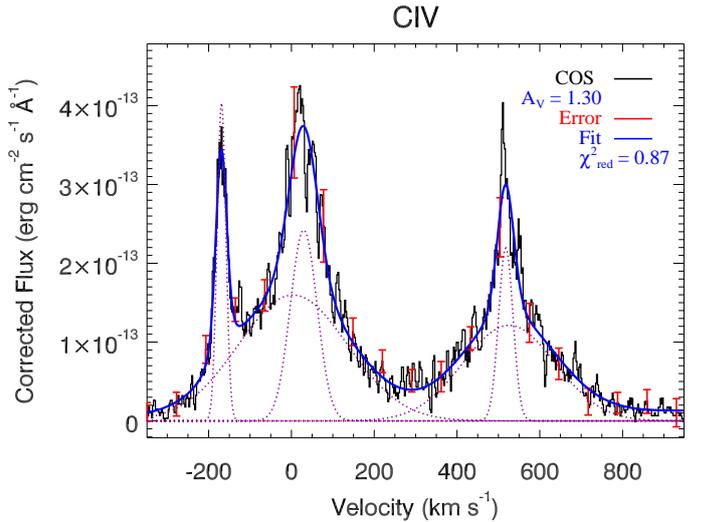}}
\caption{Line decomposition fits of the \CIV ~$\lambda$1549 doublet and the \Ht ~line. 
 The extinction corrected (A$_{\rm V}=$1.3\,mag) COS spectrum is shown with the black line, 
 whereas the fit is shown with the blue continuous line. The individual broad and narrow 
 components are shown as the dotted lines. 
   \label{lines_fits}}
\end{figure}
 
\setlength{\tabcolsep}{1pt}
\begin{table}
\caption[ ]{\label{line_fluxes} Output parameters of the fits to the \CIV ~and \Ht ~lines} 
\begin{tabular}{l|r|c|r}
\hline \hline

Line $\star$  &  V$_{\rm cen}$ ($\pm$err) & Flux ($\pm$err)        &  FWHM  ($\pm$err)  \\
      &  (km s$^{-1}$)       & ($\times10^{-14}$ erg s$^{-1}$ cm$^{-2}$) &  (km s$^{-1}$)     \\
\hline     
       &                      &                        &                   \\

 \Ht           & $-$168.9 (0.7) & ~~5.5 (0.6) & 24.9  (2.1)     \\
 \CIV(1548)-N  &     29.7 (1.7) & ~10.1 (0.9) & 75.8  (5.1)     \\
 \CIV(1548)-B  &      3.7 (4.1) &  27.2 (1.8) & 309.6 (16.5)    \\
 \CIV(1550)-N  &    517.9 (1.4) & ~~5.0 (0.6) & 41.1  (4.0)     \\
 \CIV(1550)-B  &    525.7 (2.6) &  17.5 (1.0) & 262.2 (9.7)     \\
              &                  &            &                 \\

\hline
\end{tabular}
\tablefoot{~\\
$\star$ : "N" and "B" stand for narrow and broad component, respectively}
\end{table}

A line decomposition procedure, fitting the broad and narrow components of the \CIV ~$\lambda$1549 doublet 
(see Figure~\ref{lines_fits} and Table~\ref{line_fluxes}), yields a total extinction-corrected 
\citep[A$_{\rm V}=$1.3\,mag,][]{alcala17} flux of 6.0($\pm$0.3)$\times10^{-13}$\,erg~s$^{-1}$~cm$^{-2}$. 
 Subtracting the scaled \CIV ~flux ($f_{\rm \CIV}=$ 4.6$\times10^{-13}$ erg s$^{-1}$ cm$^{-2}$) from 
 the non-accreting young star LkCa\,19 from the total flux in MY\,Lup we end up with an excess \CIV ~flux 
 of 1.4($\pm$0.3)$\times10^{-13}$\,erg~s$^{-1}$~cm$^{-2}$. 
The extinction corrected flux of the \Ht ~line is estimated at 5.5($\pm$0.6)$\times10^{-14}$\,erg~s$^{-1}$~cm$^{-2}$.
Likewise, the flux of the \CII] $\lambda$2325, measured from the extinction corrected HST NUV spectrum is 
6.0($\pm$0.5)$\times10^{-13}$\,erg~s$^{-1}$~cm$^{-2}$.

The luminosity of the lines can be computed as \Ll ~$ = 4 \pi  d^2 \cdot f_{\rm line}$, where $d=156$\,pc is the 
Gaia DR2 distance \citep[][]{gaiadr2} and $f_{\rm line}$ is the extinction-corrected flux. 
The total luminosity of the \CIV ~lines is 4.1($\pm$0.2)$\times10^{-4}$\,\Lsun, while the excess luminosity of the
line, E\CIV, with respect to the non-accreting star is 9.5($\pm$0.2)$\times10^{-5}$\,\Lsun. The luminosity of the 
\Ht ~line alone is 3.8($\pm$0.4)$\times10^{-5}$\,\Lsun, and that of the \CII] $\lambda$2325 line 
is 4.1($\pm$0.4)$\times10^{-4}$\,\Lsun.

\citet[][]{ardila13} demonstrated that the  \CIV ~line shape parameters (the velocity of the peak of the 
lines and width) do not correlate with accretion rate, but also that all accreting stars should show a broad 
component. 
The analysis of these latter authors also suggests that the accretion process first generates the broad component, with the 
narrow component becoming increasingly important at larger accretion rates. For low accretion rates 
($<$4$\times10^{-9}$\,\Msun yr$^{-1}$), the average narrow-component contribution to the luminosity is 
about 20\%, while for high accretion rates it is about 40\% on average. 
The decomposition procedure of the \CIV ~lines in MY\,Lup provides velocities and FWHM for the components  
(Table~\ref{line_fluxes}) that are more consistent with those of accreting objects than with those of 
non-accreting stars. Also, the scale factor from the 1550\,\AA ~to the 1548\,\AA ~\CIV ~line of MY\,Lup 
is 1.45, as in many accreting objects.
The contribution of the narrow components in MY\,Lup to the total luminosity is 25\%, that is, just above the 
limit for low accretors and is consistent with the narrow component contribution in other accreting objects
with similar accretion rates to those of AA\,Tau and IP\,Tau \citep[see Figure~8 in][]{ardila13}. This suggests that
the \CIV ~doublet in MY\,Lup is produced by accretion.

\subsection{Accretion properties}
According to the $L_{\rm \CIV}$ versus \Lacc ~relationships by \citet[][their Figure~10]{ingleby11} and 
\citet[][]{yang12} the \CIV ~luminosity of 4.1$\times10^{-4}$\,\Lsun ~in MY\,Lup would imply an 
accretion luminosity of $\sim$2$\times10^{-1}$\Lsun, that is, more than an order of magnitude higher 
than the upper limit derived from the optical data. 
We note that the $L$(\CII])  versus \Lacc ~relationship derived by \citet[][their Eq. 15]{ingleby13}
and the \CII] $\lambda$2325 ~luminosity estimated in the previous section yield 
\Lacc=1$\times10^{-1}$\Lsun. Such \Lacc ~values would be consistent with those
of accreting YSOs in Lupus and yield a mass accretion rate  \Macc ~$\sim$ (0.5--1)$\times10^{-8}$\,\Msun yr$^{-1}$.  
The latter  calculated as 

{\setlength{\mathindent}{0pt}
\begin{equation}
\label{Macc}
\dot{M}_{acc} = ( 1 - \frac{R_{\star}}{R_{\rm in}} )^{-1} ~ \frac{L_{acc} R_{\star}}{G M_{\star}}
 \approx 1.25 ~ \frac{L_{acc} R_{\star}}{G M_{\star}} 
,\end{equation}

assuming $\frac{R_{\star}}{R_{\rm in}}=\frac{1}{5}$,  
where $R_{\star}$ and $R_{\rm in}$ are the YSO radius and inner-disc radius, respectively
\citep[see][]{gullbring98, hart98}, and using the stellar parameters reported in \citet[][]{alcala17}.
A similar  value of  \Macc$=$8$\times10^{-9}$\,\Msun yr$^{-1}$ is derived when using the 
relationship between \CIV ~luminosity and mass accretion rate provided by \citet[][]{ardila13}, that is, 
$\log$\,\Macc$=$0.8$\times$$\log$\,$L_{\rm \CIV}-5.4$.

It is important to stress that the relationships between \CIV-luminosity and accretion properties show 
a significant scatter, mainly because of a combination of non-simultaneity and extinction errors.
Considering the total \CIV ~luminosity, and taking into account the scatter in the \citet[][]{ingleby11} 
and \citet[][]{yang12} relationships, we estimate the accretion luminosity to be in the range 
(0.9-4.0)$\times10^{-1}$\,\Lsun , which gives an accretion rate in the range (0.4--1.8)$\times10^{-8}$\,\Msun yr$^{-1}$, 
implying a range of about 0.65\,dex in $\log$\Macc. 
Similarly, when using the \citet[][]{ardila13} relationship and taking into account 
uncertainties in the component subtraction procedure and extinction, we estimate the accretion rate to be 
in the range (0.5--2.5)$\times10^{-8}$\,\Msun yr$^{-1}$, meaning a range of about 0.7\,dex in $\log$\Macc.
Although \citet[][]{ardila13} find no significant difference when using all their data or only simultaneous 
data to derive their \Macc-$L_{\rm \CIV}$ relationship, they conclude that the exact relationship remains 
uncertain. 

Using Eq.\,2 by \citet[][]{johns-krull00}, which considers correction for the presence of emission from the 
transition region, and the excess luminosity E\CIV ~calculated in the previous section, we estimate a 
mass accretion rate of 2.4$\times10^{-8}$\,\Msun yr$^{-1}$, that is, a value in the high end of the ranges quoted 
above. Considering errors and the scatter of the relationships, the \Macc ~values are in fair agreement. 
The derived \Macc ~for MY\,Lup covers a range of values accounting for the spread in possible contribution 
of the transition region.

Another important source of error is represented by the uncertainty in extinction, which in the FUV is 
a factor of about 2.6 higher than in the optical. We estimated an uncertainty in A$_{\rm V}$ of about 0.4--0.5\,mag 
\citep[][]{alcala14, alcala17}, which translates to a $\sim$0.4\,dex error in $\log$$L_{\rm \CIV}$. 
According to the $L_{\rm \CIV}$ versus \Lacc ~relationships used above, this implies \Lacc ~in the 
range (0.7--4.0)$\times10^{-1}$\,\Lsun. Thus, the range of \Macc ~due to the effect of the uncertainty 
in A$_{\rm V}$ is estimated to be (0.3--1.8)$\times10^{-8}$\,\Msun yr$^{-1}$.

We conclude that a reasonable mass accretion rate for MY\,Lup is \Macc ~$\sim$ 1($^{+1.5}_{-0.5}$)$\times10^{-8}$\,\Msun yr$^{-1}$. 
Beyond the relatively large uncertainty in \Macc, the luminosity of the \CIV ~lines suggests similar \Macc 
~values as those of bona fide accreting YSOs with comparable mass. This can be seen in Figure~\ref{Macc_Mstar} 
in Appendix~\ref{Macc_Mstar_rev}, where the stellar mass and accretion rates of the Lupus sample are revisited 
using the distances drawn from Gaia DR2. It is worth  noting that the updated value for the mass accretion rate 
is also consistent with the more luminous slab model described in Section~\ref{continuum_emiss}.

\section{Discussion}
\label{discus}

\subsection{The X-Shooter--COS-STIS discrepancy on \Lacc}
Among the several objects where HST and X-Shooter data exist, MY\,Lup is the only star fo which
a discrepancy has been observed so far. In fact, the several HST-based \Macc\  values found by \citet[][]{ingleby13} 
and those found by \citet[][]{manara14} are in good agreement. Also, an additional data set of the same observing 
program from which the MY\,Lup HST data are taken show good agreement between the HST and X-Shooter \Macc\  
values for several stars  \citep[][]{Arulanantham_prep}. Therefore, MY\,Lup seems to be the only real outlier so far.
The emission lines, and substantial FUV excess emission detected in the HST spectra of MY\,Lup strongly 
suggest that the star is an actively accreting object, whereas the X-Shooter data show weak or no accretion. 
MY\,Lup is peculiar in several respects. Besides being one of the most massive T Tauri stars in Lupus, it is an 
almost edge-on disc. Hence, both the effects of low contrast in photospheric versus continuum excess emission 
and the occultation of some accretion diagnostics may play different roles in the X-Shooter accretion 
diagnostics and in those of HST. 
In principle, the accretion flows might be obscured by the highly inclined disc hampering the detection of 
excess emission in the X-Shooter wavelength range, but such an effect would be much stronger for the \CIV ~lines. 
We note however that \citet[][]{ardila13} find no correlation between inclination and the \CIV ~luminosity
of accretion rate, hence conclude that there is no evidence that the disc or the accretion flow
are obscuring the accretion diagnostics. In their interpretation, the \CIV ~UV lines are not 
emitted from a particular place, but cover the whole star. 
On one hand, very weak or no accretion would imply very little gas content in the disc, consistent with the 
finding of a rather low gas-to-dust mass ratio from faint CO isotopologue ALMA observations by \citet[][]{miotello17}. 
Nonetheless, we highlight the fact that these authors also discuss the possibility of a normal gas-to-dust ratio but a disc highly 
depleted in CO, yielding artificially low gas mass estimates. On the other hand, the HST-UV tracers seem to 
point toward accretion more typical of Class~II discs and of some transitional discs in Lupus and 
Chamaeleon~I \citep[][]{alcala17, manara17}.

As in previous studies \citep[c.f.][]{calvet04, ingleby11, manara14}, this further confirms that FUV measurements 
represent a reliable way to establish the presence of accretion in low-accreting, early-type, intermediate-mass T~Tauri stars. It is also worth stressing that the width of the \Ha ~line in this type of object may 
already suggest active accretion \citep[Romero et al. 2012 and see also the case of SR\,21 in][]{manara14}. The
complex  \Ha ~profile, which is consistent with the model prediction by \citet[][see their Figure~13]{kurosawa06} 
of a high-inclination angle, also supports the conclusion of significant UV excess emission due to accretion, 
and of a complex geometry of the infalling streams. 

\subsubsection{Possible stellar flare ?}
One question that may arise is whether  some of the emission lines observed in the COS data, 
in particular the \Ht ~lines, could be produced by a strong flare and not by accretion. 
As discussed in \citet[][]{hoadley15} and \citet[][and references therein]{arulanantham18}, the \Ht 
~lines are formed by fluorescence of Ly-$\alpha$ photons. 
In low-mass stars, just the chromospheric Ly-$\alpha$  emission alone can generate detectable \Ht ~fluorescence 
from the stellar atmosphere \citep[][]{kruczek17}. However, a much more intense Ly-$\alpha$ pumping would be 
needed to explain the bright \Ht ~emission that we see at the distance of 156\,pc of MY\,Lup. 
In the Sun and cool dwarfs, the \Ht ~lines are produced by narrow Ly-alpha, while accretors have broad Ly-$\alpha$ 
lines.

One argument against the flare scenario is that the COS G130M and G160M data were not acquired simultaneously; 
both show the same level of \Ht ~emission, as do the G140L data.  
Furthermore, COS data is acquired in time-tag mode and light curves can be created as a function of wavelength.
Figure~\ref{UV_lightcurve} shows light curves for the \Ht ~ and \CIV ~ lines. The light curves of the G160M data 
did not show any signs of significant  temporal variability. 
Therefore, we conclude that the emission lines observed in the COS data are not related to flare events, but that 
active accretion is the explanation of the UV emission.

\begin{figure}[h]
\resizebox{1.0\hsize}{!}{\includegraphics[bb=100 70 750 450]{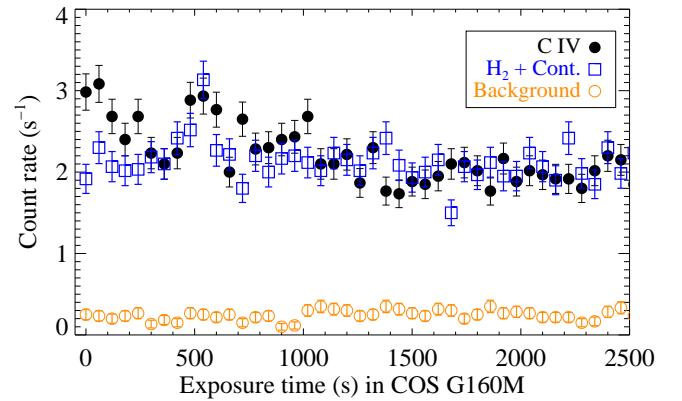}}
\caption{Far-ultraviolet light curves for a few relevant lines related to a possible flare and Ly-$\alpha$-driven fluorescence.
The \CIV ~region was integrated over 1544-1556\AA ~and the \Ht ~region was integrated over 
1488-1526\AA.  The total integration time was 2592s and the temporal cadence in the plots is 60s per time step. 
   \label{UV_lightcurve}}
\end{figure}

\subsubsection{Variability}
MY\,Lup is a variable star by definition. Time series $BVRI$ 
photometry gathered from the AAVSOnet Epoch Photometry Database\footnote{https://www.aavso.org/aavsonet-epoch-photometry-database} 
shows that the maximum amplitude of variations in the $V$-band is on the order of 0.6\,mag, with the 
tendency for the source to become redder when fainter. The latter is consistent with a variable circumstellar 
extinction due to the edge-on geometry. 
We note however that the long-term light curves from the Digital Access to a Sky Century @ Harvard (DASCH)\footnote{http://dasch.rc.fas.harvard.edu/lightcurve.php} 
suggest that events of up to 1.5\,mag in extinction may occur in MY\,Lup. 
In any case, the comparison in Figure~\ref{spec_uv_x_shoot} shows that the flux of the X-Shooter spectrum 
and the STIS spectrum are not significantly different, despite being acquired more than 2 years apart. 
The very small slope difference in the reddest spectral range can indeed be explained by an extinction change 
$\Delta$A$_{\rm V}<$0.2\,mag observed in the AAVSOnet data. Such variability would not explain an order of 
magnitude difference in the \Lacc ~estimates. 

The comparison in Figure~\ref{spec_uv_x_shoot} does not exclude, however, that larger variations 
in the UV, mostly unnoticed in the optical and due to variable accretion, may occur in other epochs. 
MY\,Lup was detected by GALEX in the NUV only, with an AB magnitude of 17.49\,mag.
A standard IDL software that converts spectra directly into a GALEX AB magnitude applied to our STIS
spectrum of MY\,Lup yields an AB magnitude of 19.40\,mag., suggesting a factor of 5--6 variability 
in the NUV flux, presumably related to accretion luminosity variability and/or inner disc occultation.  
Variability by factors of approximately five in FUV flux from accreting YSOs have previously been detected with
HST, and these tend to be sources with variable disc attenuation like AA\,Tau, RW\,Aur, and T\,Cha
\citep[see for instance][for AA\,Tau]{schneider15}. High variability has also been observed in 
multi-epoch UV HST observations of accreting low-mass stars \citep[][]{robinson_espaillat19}. 
Thus, the comparison with GALEX shows that variations in the UV in MY\,Lup are in line 
with those observed in other accreting low-mass stars. 
Objects like FUOrs and EXOrs may exhibit much higher levels of variability, but most of these objects are 
more embedded and younger than T~Tauri stars, and in the above-mentioned photometric data of MY\,Lup there 
is no evidence for a FUOri/EXLup type behaviour. We also stress that, except for the changes seen in the 
\Ha ~profile (see Figure~\ref{Ha_var}), no significant variability is observed among the two X-Shooter 
spectra available so far, either in flux or slope of the spectra. 
    
\subsubsection{Anomalous extinction}
Other relevant effects that may have an important impact on the flux measurements of the continuum 
and lines at different wavelengths are represented by a possible deviation from a normal extinction law. 
Regarding anomalous extinction, and especially for high-inclination objects, 
if one exchanged part of the absorbing material that has R$_{\rm V}=$3.1 with material that has for example
R$_{\rm V}=$4.0, this could lead to an underestimation of the \CIV ~fluxes because the FUV reddening 
correction is smaller for higher R$_{\rm V}$. However, this change might go unnoticed in the optical 
when only part of the absorber is effected. Recent SPHERE observations in scattered light provide
evidence of spatial segregation of dust grains and different (small/large) grain population and distributions
in SST\,c2dJ160830.7-382827 \citep[][]{villenave19}. This object is the other transitional 
disc found to show weak or no accretion based on the X-Shooter data (see Section~\ref{intro}) and is
very similar in many respects to MY\,Lup. However, even assuming R$_{\rm V}=$5.5 would reduce the value 
of the extinction-corrected flux of the \CIV ~lines by a factor of about four, but the resulting \Macc 
~would still be higher than the X-Shooter estimate by about 0.8\,dex in $\log$\Macc.

\subsubsection{Other effects}
The effects of scattering in the FUV may also be important. For instance, repeated observations 
of the T~Tauri star AA\,Tau have shown that the \CIV ~fluxes did not decrease as much as expected 
during the dim state \citep[see details in][]{schneider15}.

\subsubsection{Excess emission in the FUV}
\label{FUV_excess}
Previous works \citep[][]{calvet04, ingleby11, manara14} have pointed out that the excess emission for 
intermediate-mass stars is hard to detect at wavelengths longer than 3000\AA ~because the temperatures 
of the stellar photosphere and the regions of the photosphere heated by the accretion are similar. 
MY\,Lup is one of the hottest (K0-type) T~Tauri stars in Lupus, and therefore the contrast between continuum excess 
emission with respect to the photospheric+chromospheric emission, which reduces in the objects with spectral 
type earlier than about K3, may hamper detection of excess emission in the X-Shooter wavelength range.  
The  excess FUV continuum emission in MY\,Lup is evident, and is present in almost all accreting sources, 
as shown in \citet[][]{france17}. This latter study of a sample of accreting T~Tauri stars showed a strong correlation 
between the FUV continuum luminosity and the total luminosity of the \CIV ~lines 
(see their Figure~6)\footnote{We note that this is the total integrated luminosity of the \CIV ~lines, 
not only the one corresponding to the excess emission with respect to the non-accreting star.}. 
We estimated the reddening-corrected FUV continuum luminosity in MY\,Lup to be $\sim$3.6$\times$10$^{30}$\,erg~s$^{-1}$,
while the total luminosity of the \CIV ~lines is $\sim$1.6$\times$10$^{30}$\,erg~s$^{-1}$, which places 
MY\,Lup in very good agreement on that correlation. In addition, the "1600\AA ~Bump" observed in almost 
all the accreting objects investigated in \citet[][]{ingleby09}, \citet[][]{france17}, and \citet[][]{espaillat19}, 
is also present in MY\,Lup (see Figure~\ref{MYLup_LkCa19}), further putting it in the class of gas-rich, accreting 
objects.

\subsection{The MY\,Lup disc} 
 The age estimate of $\sim$17\,Myr for MY\,Lup \citep[][]{alcala17,frasca17} is much higher than the 
 average for the Lupus members, which is in principle consistent with a more  evolved  disc. 
 \citet[][]{baraffe10} concluded that episodic strong accretion during PMS evolution of low-mass stars 
 may produce objects with smaller radius, higher central temperature, and lower luminosity compared 
 to the non-accreting counterparts of the same mass and age, resulting in low-luminosity objects. Also,
 \citet[][]{kunimoto17} and \citet[][]{baraffe17} conclude that accretion history may affect PMS evolution 
 leading to a luminosity spread in the HR diagram.
 The age of MY\,Lup may however be overestimated  because the highly inclined disc partially obscures 
 the stellar photosphere. The star may appear underluminous on the HR diagram if the entire stellar surface 
 is not subject to the same extinction and part of it is almost invisible. This effect, possibly due to a warped 
 inner dusty disc, has been observed in other objects like V\,354\,Mon \citep[see][]{schneider18}. 
 Thus, MY\,Lup may indeed be much younger than what its position on the HR diagram would suggest.
 From the gravity $\log g=3.72\pm0.18$ derived by \citet[][]{frasca17}, and assuming the mass  of 1.1\Msun, 
 we estimate a radius of 2.2\Rsun ~for the star, and therefore a luminosity of 3.14\Lsun, which is a factor of about 
 four higher, placing MY\,Lup on the HR diagram in good agreement with the other Lupus members. 
 This would imply a \Macc$=$1.6$\times10^{-8}$\,\Msun yr$^{-1}$.
 This \Macc ~value estimated from the HST-UV data is indeed similar to the average derived for Lupus 
 Class~II discs. The fact that MY\,Lup shows up as a Class~II disc for the gas properties 
 is in line with other results from X-Shooter \citep[e.g.][]{manara14} and HST observations
 \citep[see the case of RY\,Lup in][but also other objects in Hoadley et al. 2015]{arulanantham18}, 
 and in many cases the gas is clearly present in the large dust cavities detected by ALMA 
 \citep[][and references therein]{miotello17,vandermarel18}. We also note that recent studies 
 \citep[][]{manara14, alcala14, manara17a, alcala17, vandermarel18} find no substantial differences 
 among the stellar and accretion properties of both primordial and transition discs 
 \citep[but see also][for a different result]{najita07, najita15}, with the exception 
 that the dusty transition discs with large cavities ($r_{cav} >$15\,AU) in Lupus tend to be more 
 massive than the Class~II disc systems \citep[][]{vandermarel18}.
 
 Combined studies of SEDs and millimetre data \citep[][]{owen_clarke12, ercolanopascucci17} suggested 
 two populations of transition discs: those with low disc masses and low or non-detectable accretion, in 
 which the inner cavities are mainly due to photoevaporation, and those with large disc masses and high 
 accretion rates, where the inner cavities are most likely created by giant planet formation. 
 Also, the \citet[][]{vandermarel18} study suggested two different evolutionary paths for the evolution of 
 protoplanetary discs: small, low-mass discs which dissipate slowly without forming large dust cavities, 
 and massive, large discs that go through a transition phase of disc clearing by giant planets. 
 
 The analysis by \citet[][]{vandermarel18} suggested a large inner dust cavity of 25\,AU in 
 the MY\,Lup disc. However, the continuum visibilities in this study show only a 
 tentative cavity at best, and the  DSHARP results of \citet[][]{huang18} show a rather centrally 
 peaked continuum emission. Therefore, a particularly deep and large dust cavity in the inner disc 
 region is not evident, casting doubts on the classification as a TD.
 Moreover, given the high mass accretion rate of  1$\times$10$^{-8}$\,\Msun yr$^{-1}$ measured from 
 the UV-HST data, a large dusty cavity would not be easily explained in terms of photoevaporation models  
 \citep[see Figure~9 in][]{vandermarel18}. Some X-ray photoevaporation 
 models might explain large cavities of accreting YSOs by imposing very low C and O disc abundances 
 \citep{ercolano17}. This would be in principle consistent with the \citet[][]{miotello17} scenario of 
 a disc highly depleted in CO, artificially leading to low gas-mass estimates. Yet, the estimated 
 $\log$\Macc~ value of $\sim -$8.0 for MY\,Lup is still high to be compatible with those models. 
 Therefore, if there is an inner cavity in the disc of MY\,Lup, it may have been caused by giant 
 planet formation.
 
 We note that the SPHERE observations by \citet[][]{avenhaus18} detected a ring in scattered light 
 at a radius of about 120\,AU from the star. Also, the annular substructures observed by DSHARP 
 \citep[][]{huang18} are possible features reproduced in simulations of planet--disc interactions. 
 Nevertheless, annular structures may not necessarily be the product of planet--disc interactions 
 \citep[see introduction in][for other possibilities]{huang18}. 
 Analysis of the ALMA and SPHERE data of MY\,Lup by C. Walsh (private communication) shows 
 that the disc is rather flared, and the gaps and rings seen in the sub-millimetre DSHARP data are relatively 
 shallow. Further detailed radiative transfer modelling of the disc using the data gathered within 
 DSHARP \citep[][]{andrews18}, as well as future direct imaging and observations at higher angular 
 and spectral resolution of molecular lines, will provide additional information for determining 
 the  origin of the disc substructures in MY\,Lup.

\section{Conclusions}
\label{conclu}

We report new HST-COS and -STIS observations of MY\,Lup, found to be a very weak or even negligible 
accretor based on previous X-Shooter observations. The UV-HST spectra show many emission lines, as well
as strong continuum excess emission in the FUV, ascribable to accretion. This confirms that MY\,Lup is actively 
accreting mass at a rate of \Macc$=$1($^{+1.5}_{-0.5}$)$\times10^{-8}$\,\Msun yr$^{-1}$, that is, a value similar to 
the average derived for Lupus Class~II discs, and more than an order of magnitude above the previous 
X-Shooter estimate. MY\,Lup is the only case so far for which a significant difference between the HST
and X-Shooter \Macc ~estimates  exists that is not ascribable to variability.
Among the several possible reasons for the X-Shooter-HST discrepancy are anomalous extinction 
and effects of scattering in the FUV. Indeed, MY Lup is peculiar in many respects: it is an almost edge-on disc and 
one of the hottest (K0-type) T~Tauri stars 
in Lupus. Hence, detection of excess emission in the X-Shooter wavelength range may be hampered both by occultation 
effects from optically thick disc material along the line of sight and by the low contrast between continuum excess 
and photospheric+chromospheric emission expected for early-type ($<$ K3) objects.
The high FUV continuum luminosity of $\sim$3.6$\times$10$^{30}$\,erg~s$^{-1}$, 
and the total luminosity of the \CIV ~lines of $\sim$1.6$\times$10$^{30}$\,erg~s$^{-1}$, as well as the 
strong \Ht ~fluorescence and 1600\AA ~Bump, place MY\,Lup in the class of gas-rich accreting objects. 
The mass accretion rate and a previously suggested large cavity of the dusty disc cannot be easily 
explained in terms of photoevaporation models. In fact, reviewing all available data suggests that no compelling 
evidence for a large dust cavity exists and we conclude that giant planet formation is simply a plausible scenario 
for the dust rings in the  disc of MY\,Lup.

\begin{acknowledgements}
We thank the referee, Greg Herczeg, for his very useful and constructive comments
which helped to improve a previous version of the paper. We thank Catherine Walsh and Nieke van der Marel 
for their very useful comments and discussions on the disc of MY\,Lup, and Antonio Frasca for help and 
discussion on the photospheric spectrum subtraction. This work has been supported by 
the project PRIN-INAF 2016 "The Cradle of Life - GENESIS-SKA (General Conditions in Early Planetary Systems 
for the rise of life with SKA)" and the PRIN-INAF-MAIN-STREAM 2017 "Protoplanetary disks seen through the
eyes of new-generation instruments". 
This research made use of the SIMBAD database, operated at the CDS (Strasbourg, France). 
CFM and AM acknowledge an ESO Fellowship. PCS acknowledges support from the Deutsches Zentrum f\"ur 
Luft- und Raumfahrt through project 50OR1706. HMG was supported by program HST-GO-15204.001, which was 
provided by NASA through a grant from the Space Telescope Science Institute, which is operated by the 
Associations of Universities for Research in Astronomy, Incorporated, under NASA contract NAS5-26555.
This work has made use of data from the European Space Agency (ESA) mission Gaia 
(https://www.cosmos.esa.int/gaia), processed by the Gaia Data Processing and Analysis Consortium 
(DPAC, https://www.cosmos.esa.int/web/gaia/dpac/ consortium). Funding for the DPAC has been provided 
by national institutions, in particular the institutions participating in the Gaia Multilateral Agreement. 
This project has received funding from the European Union's Horizon 2020 research and innovation programme 
under the Marie Sk\l{}odowska-Curie grant agreement No 823823 (DUSTBUSTERS). 
We acknowledge with thanks the variable star 
observations from the AAVSO International Database contributed by observers worldwide and used in this 
research. We also acknowledge the use of DASCH data: the DASCH project at Harvard is grateful for partial 
support from NSF grants AST-0407380, AST-0909073, and AST-1313370. This work was partly supported by the 
Deutsche Forschungs-Gemeinschaft (DFG, German Research Foundation) - Ref no. FOR 2634/1 TE 1024/1-1, 
by the DFG cluster of excellence Origin and Structure of the Universe (www.universe-cluster.de) 
\end{acknowledgements}



\begin{appendix}

\normalsize

\section{Revisited stellar and accretion properties for the Lupus sample}
\label{Macc_Mstar_rev}

Table~\ref{stell_acc_prop_gdr2} lists some of the stellar and accretion properties for the
Lupus T Tauri stars in the \citet[][]{alcala17} sample as updated using the distances 
drawn form the Gaia DR2 \citep[][]{gaiadr2}. The \Mstar ~ and \Macc ~ values are derived
using both the \citet[][]{baraffe15} and \citet[][]{siess00} evolutionary tracks, and are
respectively labelled as B15 and S00 in the table. The distances of the individual objects 
are not significantly different from those adopted in \citet[][]{alcala17} and therefore changes in 
the stellar and accretion properties are small. The main difference is that the YSOs
in the Lupus~III cloud are on the average at a distance of 159\,pc and not 200\,pc 
as previously assumed, while most of the YSOs in the other clouds are also at a distance  
of about 160\,pc and not 150\,pc. There are three problematic objects namely Sz102, 
SSTc2dJ160703.9-391112 and 2MASSJ16085373-3914367. The former is flagged with large 
parallax error, while the second has no measurement, and the third one has astrometric excess noise 
larger than 5. For these three we adopt the average value of 159\,pc.

Figure~\ref{Macc_Mstar} shows the updated $\log$\Macc--$\log$\Mstar ~plot. To generate this plot 
we use the results from the \citet{siess00} tracks for masses $\ge$ 0.1\,\Msun ~and those of 
\citet{baraffe15} for the three objects with lower values 
\citep[see discussion in Section 4.1 of][for the similarities of the tracks depending on mass]{alcala17}.

The plot shows that the accretion properties of MY\,Lup, after the new \Lacc ~ and \Macc ~estimate 
based on HST data (values in bold characters in Table~\ref{stell_acc_prop_gdr2}), are similar to 
those of transitional and Class~II discs of similar mass. 
The general conclusion on the  bi-modal behaviour of the $\log$\Macc--$\log$\Mstar ~relationship 
remains basically the same as in \citet[][]{alcala17} and \citet[][]{manara17}.
Interestingly, the transitional discs fall in general in the lower envelope of the distribution 
of points in the diagram.

\begin{figure}[h]
\resizebox{1.0\hsize}{!}{\includegraphics[bb=30 30 760 510]{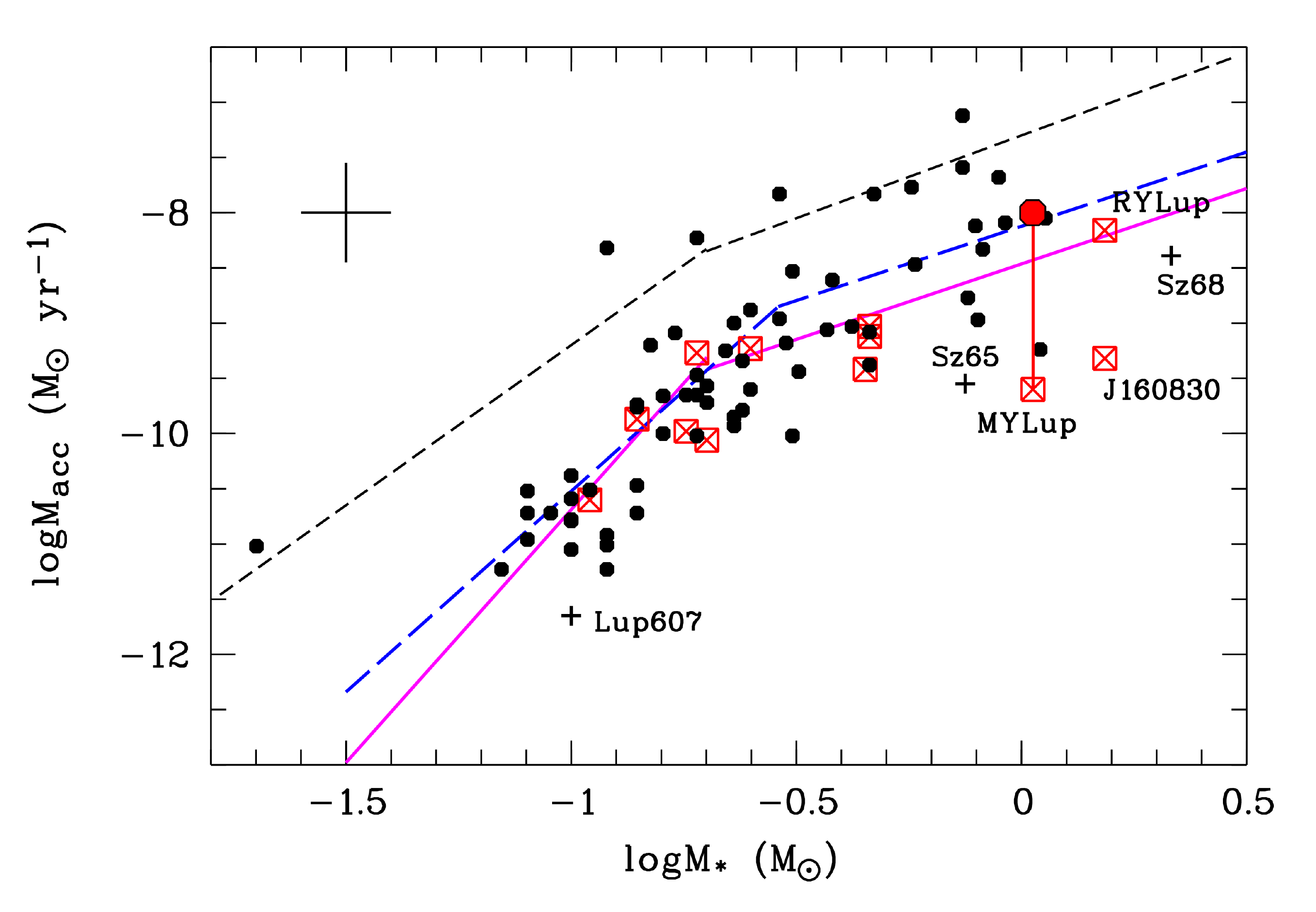}}
\caption{Mass accretion rate as a function of stellar mass for the Lupus sample of \citet[][]{alcala17}
        after revisiting the stellar and accretion parameters adopting the Gaia DR2 distances.
        The transitional discs are shown with red crossed squares. The average errors 
        in $\log$\Mstar ~and $\log$\Macc ~are shown in the upper left.  The symbols corresponding 
        to MY\,Lup, SST\,c2dJ160830.7-382827 and RY\,Lup, the latter also investigated with HST 
        \citep[][]{arulanantham18}, are labelled. The big red dot represents 
        the corrected value of \Macc ~for the former, with the vertical red line indicating 
        the corresponding vertical shift. The other three objects classified as weak or 
        negligible accretors (see Section~\ref{intro}) are plotted with plus symbols and are 
        also labelled. Sub-luminous objects are not plotted. The black dashed line shows 
        the double power law theoretically predicted by  \citet{vorobyov09}, and the 
        continuous magenta lines represent the fits to the data as in Eqs.~(4) 
        and (5) of \citet[][]{alcala17}; the long-dashed blue line shows the robust 
        double-linear fit following the prescription by \citet{manara17}. 
   \label{Macc_Mstar}}
\end{figure}

\onecolumn

\setlength{\tabcolsep}{2pt}
\begin{longtable}{l|l|c|c|c|l|c|c|c|c|c|l}
\caption[ ]{\label{stell_acc_prop_gdr2} Revisited stellar and accretion properties of Lupus YSOs.} \\
\hline \hline
        &             &       &  &    &  &  & & & & &  \\
Object  &   SpT  & $T_{\rm eff}$ & $A_{\rm V}$  & $d$  &  \Lstar   &   $\log$\Lacc  &  \Mstar (B15) &  $\log$\Macc (B15) & \Mstar (S00)  & $\log$\Macc (S00) & Notes \\
        &        & [K]           & [mag.]       & [pc] &  [\Lsun] &    [\Lsun]     &  [\Msun ]  &  [\Msun yr$^{-1}$] &  [\Msun ]     &  [\Msun yr$^{-1}$] &       \\
        &               &       &         &       &         &        &          &       &          &        &                \\                  
\hline
Sz66                    &  M3   &   3415  &  1.00 &  157 &  0.22  &  -1.76      &  0.29 &   -8.50  &   0.31 &     -8.53   &      \\
AKC2006-19              &  M5   &   3125  &  0.00 &  153 &  0.02  &  -4.08      &  0.14 &  -10.97  &   0.12 &  -10.92     &      \\
Sz69                    &  M4.5 &   3197  &  0.00 &  155 &  0.09  &  -2.77      &  0.20 &   -9.48  &   0.19 &     -9.47   &      \\
Sz71                    &  M1.5 &   3632  &  0.50 &  156 &  0.33  &  -2.17      &  0.41 &   -9.02  &   0.42 &     -9.03   &      \\
Sz72                    &  M2   &   3560  &  0.75 &  156 &  0.27  &  -1.77      &  0.37 &   -8.60  &   0.38 &     -8.61   &      \\
Sz73                    &  K7   &   4060  &  3.50 &  157 &  0.46  &  -0.96      &  0.78 &   -8.12  &   0.79 &     -8.12   &      \\
Sz74                    &  M3.5 &   3342  &  1.50 &  159 &  1.16  &  -1.45      &  0.30 &   -7.81  &   0.29 &     -7.83   &      \\  
Sz83                    &  K7   &   4060  &  0.00 &  160 &  1.49  &  -0.25      &  0.67 &   -7.08  &   0.74 &     -7.12   &      \\
Sz84                    &  M5   &   3125  &  0.00 &  153 &  0.13  &  -2.68      &  0.17 &   -9.23  &   0.19 &     -9.27   &  td  \\
Sz130                   &  M2   &   3560  &  0.00 &  160 &  0.18  &  -2.14      &  0.39 &   -9.09  &   0.37 &     -9.06   &      \\
Sz88A                   &  M0   &   3850  &  0.25 &  158 &  0.31  &  -1.40      &  0.61 &   -8.49  &   0.58 &     -8.47   &      \\
Sz88B                   &  M4.5 &   3197  &  0.00 &  159 &  0.07  &  -3.30      &  0.20 &  -10.05  &   0.19 &  -10.02     &      \\
Sz91                    &  M1   &   3705  &  1.20 &  159 &  0.20  &  -2.00      &  0.51 &   -9.07  &   0.46 &     -9.03   &  td  \\
Lup713                  &  M5.5 &   3057  &  0.00 &  174 &  0.02  &  -3.62      &  0.11 &  -10.40  &   0.10 &  -10.38     &      \\
Lup604s                 &  M5.5 &   3057  &  0.00 &  160 &  0.04  &  -3.89      &  0.12 &  -10.56  &   0.11 &  -10.51     &      \\
Sz97                    &  M4   &   3270  &  0.00 &  158 &  0.11  &  -3.11      &  0.24 &   -9.88  &   0.23 &     -9.85   &      \\
Sz99                    &  M4   &   3270  &  0.00 &  159 &  0.05  &  -2.80      &  0.23 &   -9.73  &   0.19 &     -9.65   &      \\
Sz100                   &  M5.5 &   3057  &  0.00 &  137 &  0.08  &  -3.33      &  0.14 &   -9.87  &   0.14 &     -9.87   &  td  \\
Sz103                   &  M4   &   3270  &  0.70 &  160 &  0.12  &  -2.60      &  0.23 &   -9.33  &   0.24 &     -9.34   &      \\
Sz104                   &  M5   &   3125  &  0.00 &  166 &  0.07  &  -3.36      &  0.16 &  -10.03  &   0.16 &  -10.00     &      \\
Lup706                  &  M7.5 &   2795  &  0.00 &  159 &  0.002 &  -5.00      &  0.06 &  -11.90  &          &           &  sl, ost  \\  
Sz106                   &  M0.5 &   3777  &  1.00 &  162 &  0.06  &  -2.68      &  0.57 &  -10.07  &   0.49 &  -10.00     &  sl \\  
Par-Lup3-3              &  M4   &   3270  &  2.20 &  159 &  0.15  &  -3.10      &  0.23 &   -9.77  &   0.24 &     -9.79   &      \\
Par-Lup3-4              &  M4.5 &   3197  &  0.00 &  151 &  0.002 &  -4.35      &  0.17 &  -11.81  &   0.16 &  -11.78     &  sl  \\  
Sz110                   &  M4   &   3270  &  0.00 &  160 &  0.18  &  -2.20      &  0.23 &   -8.84  &   0.25 &     -8.88   &      \\
Sz111                   &  M1   &   3705  &  0.00 &  158 &  0.21  &  -2.40      &  0.51 &   -9.47  &   0.45 &     -9.42   &  td  \\
Sz112                   &  M5   &   3125  &  0.00 &  160 &  0.12  &  -3.39      &  0.17 &   -9.94  &   0.18 &     -9.98   &  td  \\
Sz113                   &  M4.5 &   3197  &  1.00 &  163 &  0.04  &  -2.28      &  0.19 &   -9.12  &   0.17 &     -9.09   &      \\
2MASSJ16085953-3856275  &  M8.5 &   2600  &  0.00 &  150 &  0.01  &  -4.85      &  0.02 &  -11.02  &          &           &  ost \\
SSTc2d160901.4-392512   &  M4   &   3270  &  0.50 &  164 &  0.10  &  -3.17      &  0.23 &   -9.95  &   0.23 &     -9.93   &      \\
Sz114                   &  M4.8 &   3175  &  0.30 &  162 &  0.21  &  -2.68      &  0.19 &   -9.17  &   0.22 &     -9.25   &      \\
Sz115                   &  M4.5 &   3197  &  0.50 &  158 &  0.11  &  -2.91      &  0.20 &   -9.57  &   0.20 &     -9.57   &      \\
Lup818s                 &  M6   &   2990  &  0.00 &  157 &  0.02  &  -4.31      &  0.08 &  -10.96  &   0.10 &  -11.05     &      \\
Sz123A                  &  M1   &   3705  &  1.25 &  159 &  0.13  &  -2.00      &  0.55 &   -9.21  &   0.46 &     -9.12   &  td  \\
Sz123B                  &  M2   &   3560  &  0.00 &  159 &  0.03  &  -2.90      &  0.40 &  -10.24  &   0.32 &  -10.13     &  sl  \\  
SST-Lup3-1              &  M5   &   3125  &  0.00 &  165 &  0.04  &  -3.77      &  0.16 &  -10.53  &   0.14 &  -10.47     &      \\
Sz65                    &  K7   &   4060  &  0.60 &  155 &  0.89  &  -2.57      &  0.70 &   -9.52  &   0.75 &   -9.55     &      \\  
AKC2006-18              &  M6.5 &   2935  &  0.00 &  149 &  0.01  &  -4.60      &  0.07 &  -11.23  &          &           &  ost \\
SSTc2dJ154508.9-341734  &  M5.5 &   3060  &  5.50 &  155 &  0.06  &  -1.77      &  0.14 &   -8.36  &   0.12 &     -8.32   &      \\
Sz68                    &  K2   &   4900  &  1.00 &  154 &  5.42  &  -1.18      &       &          &   2.15 &    -8.39   &  obt \\  
SSTc2dJ154518.5-342125  &  M6.5 &   2935  &  0.00 &  152 &  0.04  &  -4.29      &  0.08 &  -10.72  &   0.10 &  -10.79     &      \\
Sz81A                   &  M4.5 &   3200  &  0.00 &  160 &  0.25  &  -2.44      &  0.19 &   -8.92  &   0.23 &     -9.00   &      \\
Sz81B                   &  M5.5 &   3060  &  0.00 &  160 &  0.12  &  -3.14      &  0.15 &   -9.61  &   0.16 &     -9.66   &      \\
Sz129                   &  K7   &   4060  &  0.90 &  162 &  0.43  &  -1.13      &  0.78 &   -8.30  &   0.82 &     -8.33   &      \\
SSTc2dJ155925.2-423507  &  M5   &   3125  &  0.00 &  147 &  0.02  &  -4.42      &  0.14 &  -11.29  &   0.12 &  -11.23     &      \\
RYLup                   &  K2   &   4900  &  0.40 &  159 &  1.87  &  -0.85      &       &          &   1.53 &    -8.16   &  td, obt  \\
SSTc2dJ160000.6-422158  &  M4.5 &   3200  &  0.00 &  161 &  0.10  &  -3.04      &  0.20 &   -9.73  &   0.20 &     -9.72   &      \\
SSTc2dJ160002.4-422216  &  M4   &   3270  &  1.40 &  164 &  0.18  &  -2.92      &  0.23 &   -9.56  &   0.25 &     -9.60   &      \\
SSTc2dJ160026.1-415356  &  M5.5 &   3060  &  0.90 &  164 &  0.08  &  -3.22      &  0.14 &   -9.76  &   0.14 &     -9.76   &      \\
MYLup                   &  K0   &   5100  &  1.30 &  157 &  0.85  & {\bf -0.65} &  1.09 & {\bf-8.01} &  1.06 & {\bf-7.99} &  td?  \\  
Sz131                   &  M3   &   3415  &  1.30 &  160 &  0.15  &  -2.34      &  0.30 &   -9.18  &   0.30 &     -9.18   &      \\
Sz133                   &  K5   &   4350  &  1.80 &  153 &  0.07  &  -1.78      &       &          &         &           &  sl, bz  \\  
SSTc2dJ160703.9-391112  &  M4.5 &   3200  &  0.60 &  159 &  0.003 &  -5.40      &       &          &   0.16 &  -12.76    &  sl, obt \\  
Sz90                    &  K7   &   4060  &  1.80 &  160 &  0.42  &  -1.79      &  0.78 &   -8.96  &   0.80 &     -8.97   &      \\
Sz95                    &  M3   &   3415  &  0.80 &  158 &  0.26  &  -2.70      &  0.29 &   -9.40  &   0.32 &     -9.44   &      \\
Sz96                    &  M1   &   3705  &  0.80 &  157 &  0.42  &  -2.51      &  0.45 &   -9.37  &   0.46 &     -9.38   &      \\
2MASSJ16081497-3857145  &  M5.5 &   3060  &  1.50 &  159 &  0.01  &  -3.60      &  0.10 &  -10.60  &   0.11 &  -10.60     &  td  \\
Sz98                    &  K7   &   4060  &  1.00 &  156 &  1.53  &  -0.71      &  0.67 &   -7.54  &   0.74 &     -7.59   &      \\
Lup607                  &  M6.5 &   2935  &  0.00 &  175 &  0.05  &  -5.02      &  0.09 &  -11.60  &   0.10 &  -11.65     &      \\  
Sz102                   &  K2   &   4900  &  0.70 &  159 &  0.01  &  -2.20      &       &          &         &           &  sl, bz  \\  
SSTc2dJ160830.7-382827  &  K2   &   4900  &  0.20 &  156 &  1.84  &  -2.02      &       &          &   1.53 &    -9.32   &  td  \\  
SSTc2dJ160836.2-392302  &  K6   &   4205  &  1.70 &  154 &  1.15  &  -1.03      &  0.83 &   -8.04  &   0.92 &     -8.09   &      \\
Sz108B                  &  M5   &   3125  &  1.60 &  169 &  0.11  &  -3.05      &  0.17 &   -9.62  &   0.18 &     -9.65   &      \\
2MASSJ16085324-3914401  &  M3   &   3415  &  1.90 &  168 &  0.21  &  -3.25      &  0.29 &  -10.00  &   0.31 &  -10.02     &      \\
2MASSJ16085373-3914367  &  M5.5 &   3060  &  4.00 &  159 &  0.00  &  -3.90      &  0.10 &  -10.94  &   0.12 &  -11.01     &      \\
2MASSJ16085529-3848481  &  M6.5 &   2935  &  0.00 &  158 &  0.05  &  -4.31      &  0.09 &  -10.72  &   0.10 &  -10.78     &      \\
SSTc2dJ160927.0-383628  &  M4.5 &   3200  &  2.20 &  159 &  0.07  &  -1.50      &  0.20 &   -8.25  &   0.19 &     -8.23   &      \\
Sz117                   &  M3.5 &   3340  &  0.50 &  159 &  0.28  &  -2.30      &  0.25 &   -8.91  &   0.29 &     -8.96   &      \\
Sz118                   &  K5   &   4350  &  1.90 &  164 &  0.72  &  -1.97      &  1.04 &   -9.21  &   1.10 &     -9.24   &      \\
2MASSJ16100133-3906449  &  M6.5 &   2935  &  1.70 &  193 &  0.19  &  -3.43      &       &          &   0.14 &    -9.74   &  obt \\
SSTc2dJ161018.6-383613  &  M5   &   3125  &  0.50 &  159 &  0.04  &  -4.00      &  0.15 &  -10.76  &   0.14 &  -10.72     &      \\
SSTc2dJ161019.8-383607  &  M6.5 &   2935  &  0.00 &  159 &  0.04  &  -4.10      &  0.08 &  -10.52  &   0.10 &  -10.59     &      \\
SSTc2dJ161029.6-392215  &  M4.5 &   3200  &  0.90 &  163 &  0.11  &  -3.38      &  0.20 &  -10.05  &   0.20 &  -10.06     &  td  \\
SSTc2dJ161243.8-381503  &  M1   &   3705  &  0.80 &  160 &  0.39  &  -2.19      &  0.45 &   -9.07  &   0.46 &     -9.08   &      \\
SSTc2dJ161344.1-373646  &  M5   &   3125  &  0.60 &  160 &  0.04  &  -2.49      &  0.16 &   -9.24  &   0.15 &     -9.20   &      \\
Sz75                    &  K6   &   4205  &  0.70 &  152 &  1.48  &  -0.69      &  0.80 &   -7.63  &   0.89 &     -7.68   &      \\
Sz76                    &  M4   &   3270  &  0.20 &  160 &  0.18  &  -2.55      &  0.23 &   -9.18  &   0.25 &     -9.23   &  td  \\
Sz77                    &  K7   &   4060  &  0.00 &  155 &  0.59  &  -1.67      &  0.75 &   -8.76  &   0.76 &     -8.77   &      \\
RXJ1556.1-3655          &  M1   &   3705  &  1.00 &  158 &  0.26  &  -0.85      &  0.49 &   -7.85  &   0.47 &     -7.83   &      \\
Sz82                    &  K5   &   4350  &  0.90 &  159 &  2.60  &  -1.05      &  0.95 &   -7.98  &   1.13 &     -8.05   &      \\
EXLup                   &  M0   &   3850  &  1.10 &  158 &  0.76  &  -0.91      &  0.53 &   -7.74  &   0.57 &     -7.77   &      \\

 & & & &  &  &  & & & & & \\
\hline

\end{longtable}
\tablefoot{\\
td: transitional disc \\
sl: sub-luminous YSO \\
bz: sub-luminous object falling below the zero-age main sequence on the HR diagram\\
ost: outside mass range of the S00 tracks\\
obt: outside mass range of the B15 tracks\\
}

\end{appendix}

\end{document}